\documentclass[seceq]{ptptex}

\usepackage{graphicx}
\usepackage{wrapft}

\newcommand{\bk}[1]{\left( #1 \right)}

\title{
Induced Order in Nonequivalent Two-Leg Hubbard Ladder
}

\author{
Hiroyuki \textsc{Yoshizumi}\footnote{E-mail: ydsumi@yukawa.kyoto-u.ac.jp},
Takami \textsc{Tohyama} and Takao \textsc{Morinari}
}

\inst{
Yukawa Institute for Theoretical Physics, Kyoto University,\\
Kyoto 606-8502, Japan
}

\abst{
Motivated by the presence of different orders in multilayered high-temperature superconductors, we examine a model consisting of nonequivalent two Hubbard chains coupled by interchain hopping by using the density-matrix renormalization group (DMRG) and a mean-field theory. As an example, we consider a system with noninteracting chain without order and a Hubbard chain with strong antiferromagnetic (AF) spin-density-wave correlation. We find that the magnitude of the interchain hopping controls the strength of induced AF correlation as well as that of original one. It is also found that the induced AF correlation decreases with increasing the magnitude of the original correlation. Implications to the multilayered system are discussed.
}

\begin{document}

\maketitle

\section{Introduction}

Recently the coexistence of antiferromagnetism and superconductivity has been reported in multilayered high-$T_c$ cuprate superconductors from nuclear magnetic resonance experiments.~\cite{multihtsc,multihtsc2} In the superconductors, there are different layers whose carrier concentrations are unequal. When the concentration is small (large), antiferromagnetic (superconducting) order is dominating. At intermediate concentration region, the coexistence of both the orders is realized. The coexistence itself has theoretically been examined by using models with strong correlation like a $t$-$J$ model within an isolated single layer.~\cite{Ogata08} However, in addition to the possible coexistence in the single layer, we need to clarify a role of induced order from neighboring ordered layers due to proximity effect, when we analyze the experimental results for the coexistence.  

In order to exploit the induced order, we consider a simple model that can describe the proximity effect: a two-leg ladder system with different interaction parameters in each leg and coupled with one-particle electron hopping interaction. In this paper, we examine a two-leg ladder consisting of a noninteracting chain and of a Hubbard chain where antiferromagnetic (AF) spin correlation is dominating. It is noted that, though two-leg ladder systems with equivalent chains have been studied intensively by both analytical~\cite{eqladderana} and numerical~\cite{eqladdernum} approaches, nonequivalent ladder systems have not been investigated as far as we know. For this purpose, we employ density-matrix renormalization group (DMRG) technique to calculate the distance dependence of spin-spin correlation function. We find that the magnitude of the interchain hopping controls the strength of induced AF spin correlation in the noninteracting chain as well as that of original correlation
  in the Hubbard chain. Furthermore, it is found that the induced AF correlation decreases with increasing the magnitude of the original correlation. This is reproduced by a mean-field treatment for the ladder system, indicating that these findings are independent of dimensionality.

This paper is organized as follows. In \S2, we introduce a model Hamiltonian of a two-leg ladder consisting of nonequivalent chains described by the Hubbard model and coupled with interchain hopping interaction. In \S3, we present DMRG results and discuss how the interchain hopping affects on original and induced AF spin correlations on each chain. In \S4, using an analytical calculation, we clarify a relation between the strengths of induced and original orders. \S5 is devoted to summary of this work.

\begin{figure}[tb]
\centerline{\includegraphics[width=8cm]{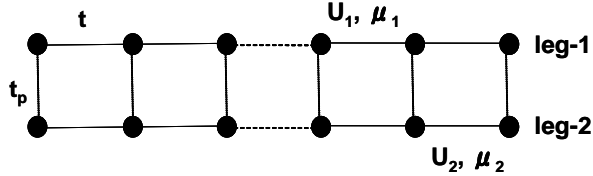}}
\caption{Schematic view of a two-leg Hubbard ladder. The hopping parameter $t$ is common to both the leg 1 and leg 2, but the on-site Coulomb interaction $U_l$ and the chemical potential $\mu_l$ depend on the $l$-th leg.}
\label{fladder}
\end{figure}

\section{Model}
We investigate a ladder system consisting of two nonequivalent Hubbard chains connected by a hopping term, as schematically shown in Fig.~\ref{fladder}. The Hamiltonian of the two-leg Hubbard ladder model is given by
\begin{equation}
\mathcal{H}=
\sum_{l=1,2}\mathcal{H}_l+\mathcal{H}_\perp
\label{TotalH}
\end{equation}
with
\begin{equation}
\mathcal{H}_l=
-t\sum_{i,\sigma}\bk{c^\dagger_{i,l,\sigma}c^{}_{i+1,l,\sigma}+\mathrm{h.c.}}+U_l\sum_in_{i,l,\uparrow}n_{i,l,\downarrow}+\mu_l \sum_i n_{i,l}
\end{equation}
and
\begin{equation}
\mathcal{H}_\perp=
-t_p\sum_{i,\sigma}\bk{c^\dagger_{i,1,\sigma}c^{}_{i,2,\sigma}+\mathrm{h.c.}},
\end{equation}
where $c^{}_{i,l,\sigma}$ ($c^\dagger_{i,l,\sigma}$) annihilates (creates) an electron of spin $\sigma$ at the $i$-th rung on leg $l$, $n_{i,l}=\sum_\sigma n_{i,l,\sigma}=\sum_\sigma c_{i,l,\sigma}^\dagger c_{i,l,\sigma}^{}$, $t$ is the intra-chain hopping parameter common to the legs, $t_p$ is the inter-chain hopping parameter, $U_l$ is the on-site Coulomb repulsion on leg $l$, and $\mu_l$ is the chemical potential that controls the charge density on each leg. Hereafter we take $t=1$.

When $t_p=0$, the ladder represents decoupled chains. The parameters in the chains determine dominant correlations among spin, charge, and superconducting orders at half filling.~\cite{1Dphase} In this study, we consider a simple case where the $l=1$ leg is noninteracting ($U_1=0$) and the $l=2$ leg has positive number of $U_2$ giving rise to the longest power-low correlation for spin channel. In the following, we call the $l=1$ leg the metal chain and the $l=2$ leg the spin-density-wave (SDW) chain. We connect the two chains by introducing $t_p$, whose value is assumed to be small, i.e., $t_p<t$, keeping the multilayer cuprates in mind. In order to escape the change of correlation functions due to charge transfer from one chain to another, the chemical potential $\mu_l$ is chosen to make each chain contain a half of total electrons, i.e., the averaged electron number in the $l$ leg is given by $\sum_i\langle n_{i,l}\rangle=N/2$, where $N$ is the number of total lattice sites
  and $\langle\cdots\rangle$ stands for the ground-state expectation value. In this study, we consider the case that the total number of electrons is equal to $N$, i.e., half filling, and take $\mu_l=U_l/2$.

Employing the finite-size algorithm of density matrix renormalization group (DMRG) method~\cite{white} for a ladder with $N=64\times2$ sites under open boundary conditions, we calculate the spin-spin correlation function for each leg, which is defined as
\begin{equation}
C_l\bk{\left|r_i-r_j\right|}=\langle S^z_l(r_i)S^z_l(r_j)\rangle,
\end{equation}
where $S^z_l(r_i)$ is the $z$ component of the spin operator at the $i$-th position $r_i$ on the leg $l$. The middle of the site $i$ and the site $j$ is chosen to be at the center of the leg to minimize size effects coming from edges of the ladder.

\section{DMRG results}

\subsection{Spin-spin correlation in the SDW chain}

\begin{figure}[tb]
\centerline{\includegraphics[width=10cm]{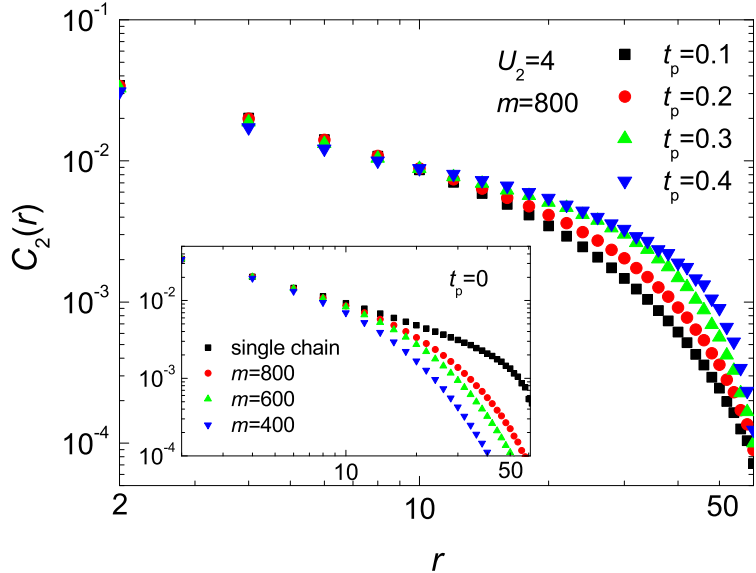}}
\caption{Spin-spin correlation function on the SDW chain $C_2(r)$ as a function of the distance $r$ between two sites for a 64$\times$2 ladder with $U_2=4$. The data for several values of the interchain hopping $t_p$ are shown. The truncation number $m=800$. Inset: $m$ dependence of $C_2(r)$ for $t_p=0$ by using the ladder system. The filled squares represent the case of a 64-site single chain.}
\label{OriginalC}
\end{figure}

We first examine spin-spin correlation in the SDW chain $C_2(r)$ (the $l=2$ leg). Figure~\ref{OriginalC} shows the correlation function for several values of the interchain hopping $t_p$ in the case of $U_2=4$. We plot only positive values of the correlation where the distance $r$ is of even number. We note that odd distances give negative values of the correlation. 

Before discussing the details of the results, we should mention convergency of the data in terms of the truncation number $m$ in DMRG. Since we treat a ladder system with a noninteracting chain and small $t_p$, we are required to use a large number of $m$. Inset in Fig.~\ref{OriginalC} shows $m$ dependence of $C_2(r)$ for the case of $t_p=0$. Since the two chains are decoupled, $C_2(r)$ must agree with that of a 64-site single chain with on-site Coulomb interaction $U=4$, which shows a linear dependence in the log-log plot for $r\lesssim 40$ (solid squares). With increasing $m$ from 400 to 800, $C_2(r)$ approaches that of the single chain. The data at short distances ($r\lesssim 10$) seems to converge in the logarithmic scale for $m=800$, while the data at long distances show slow convergencey. Such slow convergency comes from the small correlation between the two chains. The convergency in terms of $m$ becomes much better with increasing $t_p$. However, $m=800$ are not enough to get good convergency at large distances. Unfortunately $m=800$ is a maximum number we can access using our available computers. We note that CPU time for the calculation of the correlation function for a given parameter is about 52 hours by using 64 cores on FUJITSU HX600. We also note that truncation error for $m=800$ is about $10^{-4}$.

\begin{figure}[tb]
\centerline{\includegraphics[width=9cm]{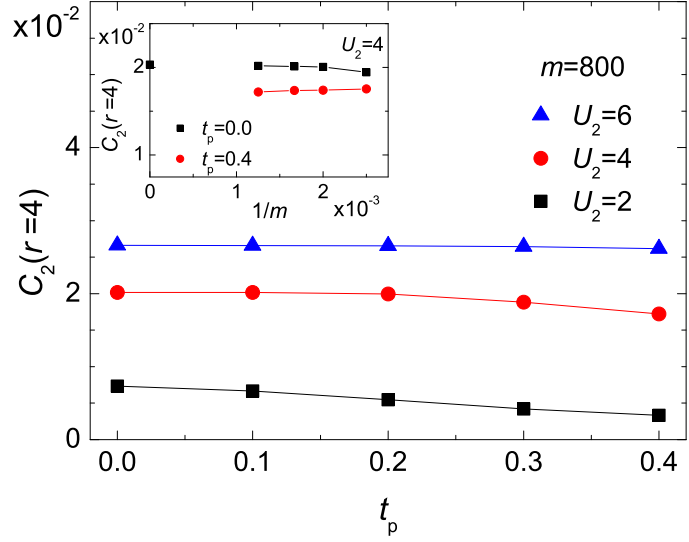}}
\caption{$t_p$ dependence of spin-spin correlation on the SDW chain at a selected distance $r=4$ for a 64$\times$2 ladder and different values of $U_2$. The truncation number $m=800$. Inset: $C_2(r=4)$ for $U_2=4$ as a function of the inverse of $m$. The data for $t_p=0$ and 0.4 are shown. The filled square at $1/m=0$ is the data obtained from a 64-site single chain.}
\label{OriginalCUdep}
\end{figure}

Keeping in mind the convergency problem mentioned above, let us discuss how $C_2(r)$ is affected by the introduction of $t_p$. For $r\lesssim 10$, $C_2(r)$ slightly decreases with increasing $t_p$. In order to clarify the effect of the on-site Coulomb interaction, we show in Fig.~\ref{OriginalCUdep}(a) $t_p$ dependence of the spin-spin correlation at $r=4$ on the SDW chain for $U_2=2$, 4, and 6. For the check of convergency with respect to $m$, we show $1/m$ dependence of $C_2(r=4)$ for $U_2=4$ in the inset. The data for $t_p=0$ seems to be safely extrapolated to the data of a single chain plotted at $1/m=0$. Similarly, the data for $t_p=0.4$ exhibits a smooth change toward $1/m=0$. These indicate good convergency for $m=800$. In the main panel of Fig.~\ref{OriginalCUdep}, $C_2(r=4)$ is suppressed with increasing $t_p$. This is easily understood as a result of mixing with the connected chain. Such suppression has been discussed in the context of the effect of apical oxygen on 
 pairing amplitude in the CuO$_2$ plane,~\cite{Mori08} though the dimensionality is higher than the present case. The suppression of $C_2(r=4)$ depends on $U_2$. With increasing $U_2$, the suppression is reduced and very small at $U_2=6$. This means that, at large $U_2$, $C_2$ hardly changes even if $t_p$ is introduced. This is also easily understood by the fact that the double occupation on the SDW chain is not allowed during the process of interchain hopping. This effect is also seen in the metal chain as a suppression of induced AF correlation, as will be discussed below. 

At large distances more than $r\sim 10$, $C_2(r)$ increases with increasing $t_p$ as shown in Fig.~\ref{OriginalC}. Since the convergency of the data at this region is questionable as mentioned above, we cannot give a definite answer to a question whether $C_2(r)$ increase with $t_p$ or not. We need larger $m$ in DMRG or analytical studies in order to clarify this question. This remains as a future work. We note that a linear behavior up to $r\sim 30$ in the log-log plot for the case of $t_p=0.4$ indicates no spin gap in the SDW chain. However, we need further study to draw a final conclusion.

\subsection{Spin-spin correlation in the metal chain}

\begin{figure}[tb]
\parbox{\halftext}{\includegraphics[width=\halftext]{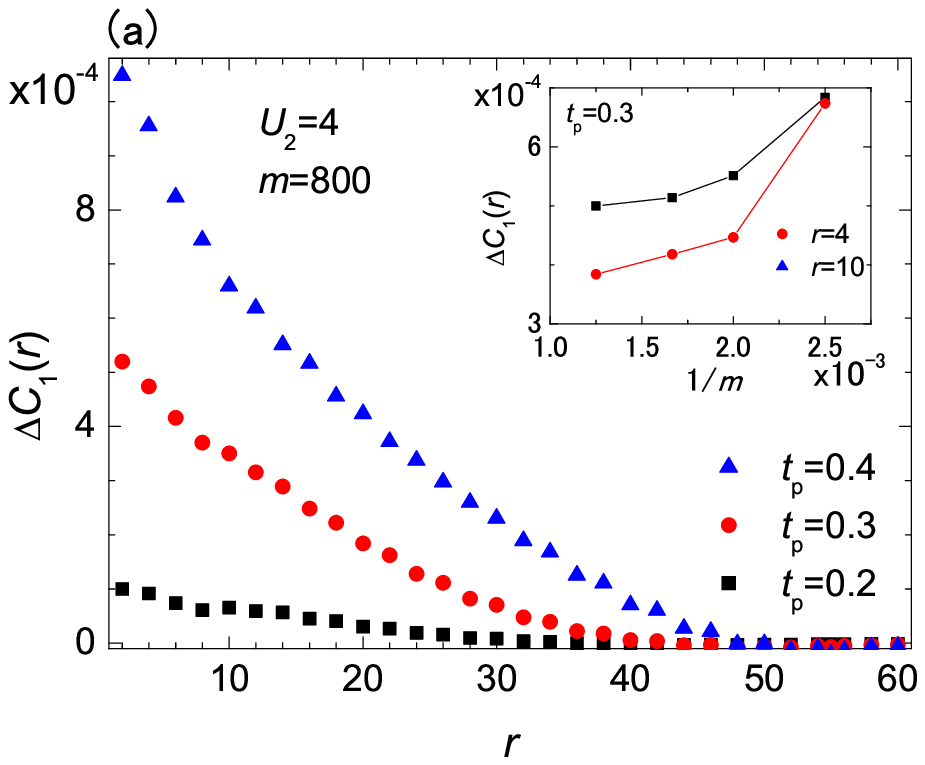}}
\hfill
\parbox{\halftext}{\includegraphics[width=\halftext]{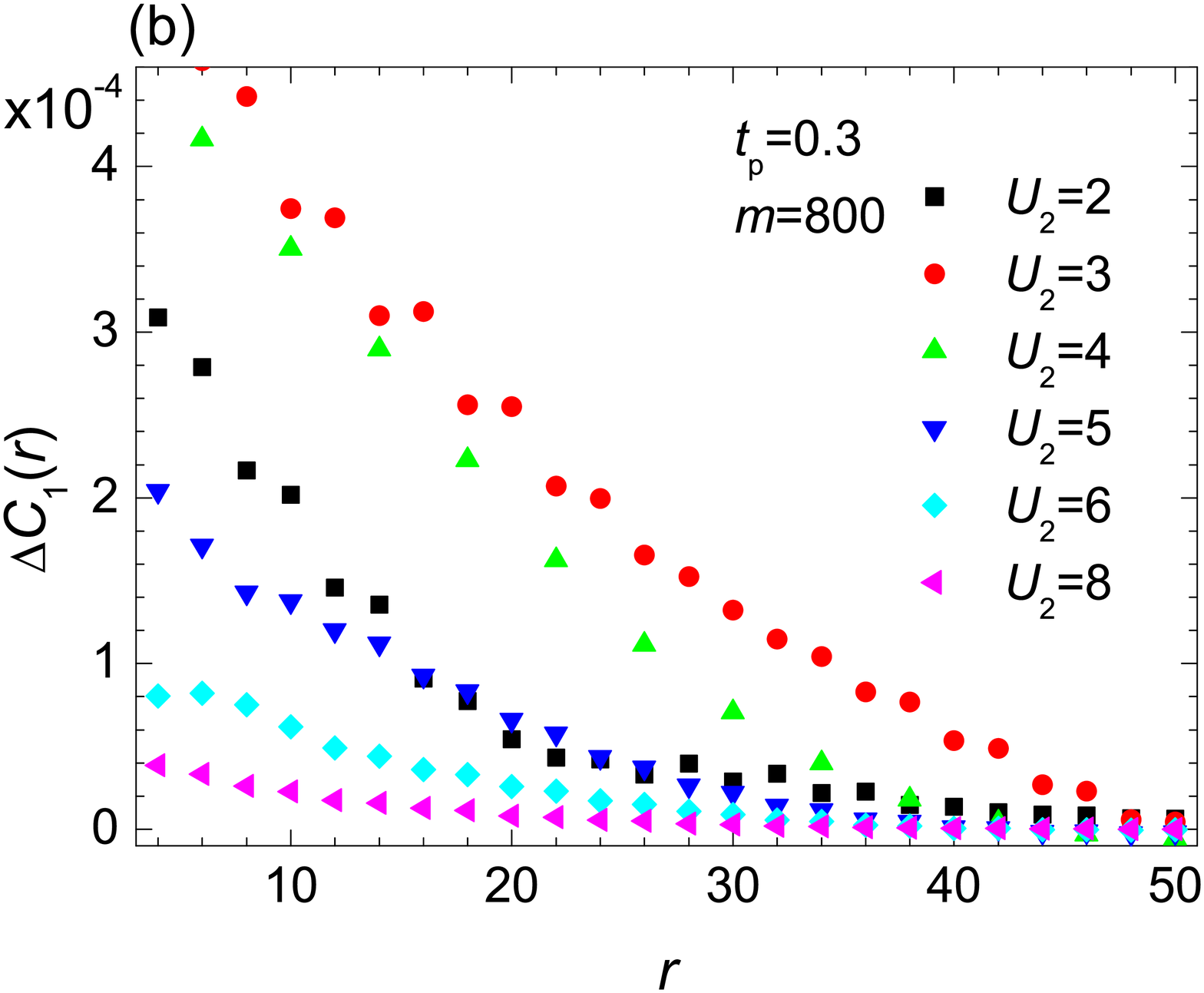}}
\caption{Induced AF spin correlation function on the metal chain $\Delta C_1(r)$ as a function of the distance $r$ between two sites for a 64$\times$2 ladder. The truncation number $m=800$. (a) $U_2=4$ and the data for several values of $t_p$ are shown. Inset: $\Delta C_1(r)$ at several distances for $t_p=0.3$ as a function of $1/m$. (b) $t_p=0.3$ and the data for several values of $U_2$ are shown.}
\label{InducedC}
\end{figure}

We show in Fig.~\ref{InducedC} induced AF spin correlation function $\Delta C_1$ in the metal chain (the $l=1$ leg), which is defined as the difference of the correlations for finite $t_p$ and $t_p=0$. The data at even distances are plotted. At $U_2=4$, the magnitude and length scale of $\Delta C_1$ increase with increasing $t_p$ as shown in Fig.~\ref{InducedC}(a). As for convergency of the data in terms of $m$, we again plot $1/m$ dependence of $\Delta C_1$ at several values of $r$ for $t_p=0.3$ in the inset of Fig.~\ref{InducedC}(a). The data gradually converges from $m=500$ ($1/m=0.002$) to $m=800$ ($1/m=0.00125$). In Fig.~\ref{InducedC}(b), we plot $U_2$ dependence of the induced AF correlation for a fixed value of $t_p=0.3$. With increasing $U_2$, $\Delta C_1$ increases and saturates near $U_2=3$. Then, $\Delta C_1$ decreases with further increasing $U_2$.

In a single Hubbard chain, the magnitude of the AF spin correlation increases with increasing $U$. The increse of $U$ also reduces charge fluctuation. This relation between AF correlation and charge fluctuation tempts us to relate the nonmonotonic $U_2$ dependence of the induced AF correlation $\Delta C_1$ to charge fluctuation in the metal chain. In order to confirm this, we have calculated the charge fluctuation on a central site  of the metal chain, defined as $(\Delta n)^2=\langle n_{i=32,l=1}^2\rangle-\langle n_{i=32,l=1}\rangle^2$ (not shown). As expected, $(\Delta n)^2$ decreases from $U_2=2$ to 3, and then increases with further increasing $U_2$, although the change of its magnitude is only 0.5\%. This implies that the induced AF spin correlation is related to charge fluctuation affected by the connected SDW chain.

\begin{figure}[tb]
\centerline{\includegraphics[width=\halftext]{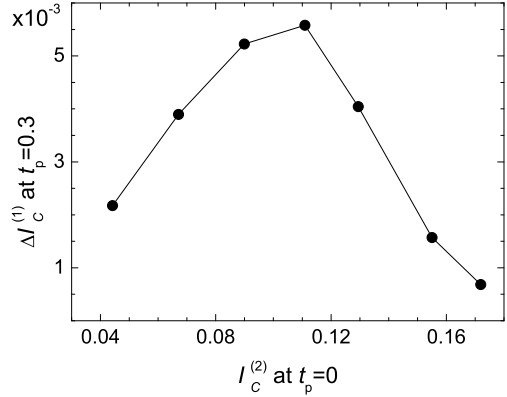}}
\caption{Relation between integrated AF spin correlation $I_C^{(2)}$ on the SDW chain at $t_p=0$ and integrated induced AF spin correlation $\Delta I_C^{(1)}$ on metal chain at $t_p=0.3$ for a 64$\times$2 ladder. The data are taken from DMRG results with $m=800$.}
\label{InducedvsOriginal}
\end{figure}

In order to make more clear the $U_2$ dependence of $\Delta C_1$, we examine a relation between the original AF spin correlation on the SDW chain at $t_p=0$ and the induced correlation at $t_p=0.3$ on the metal chain. For this purpose, we define the following integrated correlation functions: 
\begin{eqnarray}
I_C^{(2)}&=&\sum_{i-j=\mathrm{even}} C_2\bk{\left|r_i-r_j\right|},\\
\Delta I_C^{(1)}&=&\sum_{i-j=\mathrm{even}} \Delta C_1 \bk{\left|r_i-r_j\right|},
\end{eqnarray}
where $i-j$ runs over even numbers, i.e., $i$ and $j$ are on the same sublattice. Therefore, $I_C^{(2)}$ is roughly proportional to the magnitude of spin structure factor at momentum $Q=\pi$ on the SDW chain. Similarly, $\Delta I_C^{(1)}$ roughly corresponds to the induced part of the $Q=\pi$ structure factor on the metal chain. We show $\Delta I_C^{(1)}$ versus $I_C^{(2)}$ in Fig.~\ref{InducedvsOriginal}. Each data point corresponds to a given value of $U_2$. We find that the induced AF correlation shows non-monotonic dependence on the original AF correlation on the SDW chain. In particular, the induced correlation has a maximum for a certain value of the original correlation and weakens with further increasing the strength of the original correlation. This behavior seems to be counterintuitive, since induced correlation is expected to be strong if original correlation is strong. In the next section, we consider the origin of this behavior using a mean-field treatment of the
  ladder system.

\section{Mean-field analysis}

We perform a mean-field SDW approximation for the SDW chain (the $l=2$ leg). By assuming an ordered momentum $Q=\pi$ under periodic boundary condition, the corresponding SDW order parameter is given by $\Delta_2=\frac{1}{2}\langle n_{i2\uparrow}-n_{i2\downarrow}\rangle (-1)^{r_i}$, which becomes independent of site $i$. Using $\Delta_2$, we obtain the Hamiltonian of the nonequivalent two-leg Hubbard ladder:
\begin{equation}
\mathcal{H}_\mathrm{MF}=\tilde{\mathcal{H}}_1+\tilde{\mathcal{H}}_2+\mathcal{H}_\perp
\end{equation}
with
\begin{eqnarray}
\tilde{\mathcal{H}}_1&=&
\sum_{k}\tilde{a}_{k,1}^\dagger
\left(\begin{array}{cccc}
\xi_k&0&0&0\\
0&\xi_k&0&0\\
0&0&\xi_{k+Q}&0\\
0&0&0&\xi_{k+Q}
\end{array}\right)\tilde{a}_{k,1}, \\
\tilde{\mathcal{H}}_2&=&
\sum_{k}\tilde{a}_{k,2}^\dagger
\left(\begin{array}{cccc}
\xi_k&0&-U_2\Delta_2&0\\
0&\xi_k&0&U_2\Delta_2\\
-U_2\Delta_2&0&\xi_{k+Q}&0\\
0&U_2\Delta_2&0&\xi_{k+Q}
\end{array}\right)\tilde{a}_{k,2}, \\
\end{eqnarray}
and
\begin{equation}
\mathcal{H}_\perp
=-t_p\sum_{k,\sigma}\left(c^\dagger_{k,1,\sigma}c^{}_{k,2,\sigma}+c^\dagger_{k+Q,1,\sigma}c_{k+Q,2,\sigma}+\mathrm{h.c.}\right),
\end{equation}
where $\xi_k=-2t\cos k-\mu$ with the chemical potential $\mu$ and $\tilde{a}^\dagger_{k,l}= (c_{k,l,\uparrow}^\dagger,c_{k,l,\downarrow}^\dagger,c_{k+Q,l,\uparrow}^\dagger,c_{k+Q,l,\downarrow}^\dagger)$, with $c_{k,l,\sigma}^\dagger$ being the Fourier component of $c_{i,l,\sigma}^\dagger$. Integrating out the freedom of the SDW chain, the inverse of an effective Green function on the $l=1$ leg reads
\begin{eqnarray}
&&G^{-1}_{\mathrm{eff},1}\left(k,i\omega_n\right) \nonumber\\
&=&i\omega_n
-\left(\begin{array}{cccc}
B_n(k,\Delta_2)&0&-D_n(k,\Delta_2)&0 \\
0&B_n(k,\Delta_2)&0&D_n(k,\Delta_2) \\
-D_n(k,\Delta_2)&0&C_n(k,\Delta_2)&0 \\
0&D_n(k,\Delta_2)&0&C_n(k,\Delta_2)
\end{array}\right),
\label{Geff}
\end{eqnarray}
where $B_n(k,\Delta_2)=\xi_k+A_n(k,\Delta_2)\bk{i\omega_n-\xi_{k+Q}}$, $C_n(k,\Delta_2)=\xi_{k+Q}+A_n(k,\Delta_2)\bk{i\omega_n-\xi_k}$, $D_n(k,\Delta_2)=U_2\Delta_2 A_n(k,\Delta_2)$, and $A_n(k,\Delta_2)=t_p^2/\bigl[(i\omega_n-\xi_k)(i\omega_n-\xi_{k+Q})-\bk{U_2\Delta_2}^2\bigr]$, with $\omega_n$ being the Matsubara frequency

By relating the off-diagonal elements in (\ref{Geff}) to the induced order, an effective SDW order emerging on the $l=1$ leg, $\Delta^\mathrm{eff}_1$, is given by $D_n(k,\Delta_2)/U_2$. Considering electrons near the Fermi level at half-filing, i.e., $k=\pi/2$ and $\mu=0$, and assuming $t_p\ll\Delta_2$, we obtain
\begin{equation}
\Delta^\mathrm{eff}_1\approx \frac{t_p^2}{U_2^2}\frac{1}{\Delta_2},
\label{edelta2}
\end{equation}
where the induced order decreases inversely proportional to the magnitude of the original order. This behavior is qualitatively consistent with the DMRG results of the induced AF spin correlation as shown in Fig.~\ref{InducedvsOriginal}. Of course, the quantities calculated in the DMRG and mean-field calculations are different, but underlying physical origins should be the same because both the correlation and the order are related to the interplay of on-site Coulomb interaction and the spin degrees of freedom. From the mean-field result, we can obtain an intuitive picture of the decrease of the induced order at large original order. For large original order, an electron in the metal chain feels strong potential due to the order and thus the electron is hard to hop to the SDW chain. As a result, the transfer of the order to the metal chain is less effective during the second order process using $t_p$ as seen in (\ref{edelta2}). It is interesting to note that this mechanism is
  independent of dimensionality.

\section{Summary}
We have examined the spin-spin correlation function of a nonequivalent two-leg Hubbard ladder in order to exploit the induced order due to the proximity effect. One leg is a noninteracting chain and the other is a Hubbard chain with positive on-site Coulomb interaction where AF spin correlation is dominating. The two chains are coupled by interchain hopping. By employing the DMRG technique, we find that the AF correlation on the SDW chain is slightly suppressed at short distances with increasing interchain hopping. This suppression is due to the transfer of AF spin correlation to the metal chain. Then, the induced AF spin correlation in the metal chain is enhanced. Fixing $t_p$, we find that the induced correlation decreases with increasing the magnitude of the original correlation in the SDW chain. This is supported by a mean-field SDW calculation.

The model taken in this study is too simple to describe multilayered high-$T_c$ cuprates in which the coexistence of antiferromagnetic and superconducting orders has been reported. The issue of the coexistence is beyond the present study, but a possible induced order due to proximity effect can be discussed based on the present results of DMRG and mean-field calculations. The results indicate that, if the order is large in one CuO$_2$ plane, the induced order in neighboring planes should be small. This means that, if the inner CuO$_2$ plane in the multilayered cuprates is close to half filling where the magnetic moment is large, induced part of the moment in the outer plane should be small. The present results might contribute to further understanding the coexistence of different orders in the multilayered cuprates.

\section*{Acknowledgements}
This work was supported by Nanoscience Program of Next Generation Supercomputing Project, and the Grant-in-Aid for Scientific Research (18340097), that on Priority Areas ``Novel States of Matter Induced by Frustration'' (19052003), the Grant-in-Aid for the Global COE Program ``The Next Generation of Physics, Spun from Universality and Emergence'' from MEXT, the Yukawa International Program for Quark-Hadron Sciences at YITP, Kyoto University. A part of numerical calculations was performed in the supercomputing facilities in ISSP, University of Tokyo, YITP and ACCMS, Kyoto University.

%

\end{document}